# Changes of Magnetism in a Magnetic Insulator due to Proximity to a Topological Insulator

Tao Liu[1], James Kally[2], Timothy Pillsbury[2], Chuanpu Liu[1], Houchen Chang[1], Jinjun Ding[1], Yang Cheng[3], Maria Hilse[2], Roman Engel-Herbert[2], Anthony Richardella,[2] Nitin Samarth[2], and Mingzhong Wu[1]†

[1]*Department of Physics, Colorado State University, Fort Collins, Colorado 80523, USA*
[2]*Materials Research Institute, Pennsylvania State University, University Park, Pennsylvania 16802, USA*
[3]*Department of Physics, The Ohio State University, Columbus, Ohio 43210, USA*

This letter reports the modification of magnetism in a magnetic insulator $Y_3Fe_5O_{12}$ thin film by topological surface states (TSS) in an adjacent topological insulator $Bi_2Se_3$ thin film. Ferromagnetic resonance measurements show that the TSS in $Bi_2Se_3$ produces a perpendicular magnetic anisotropy, results in a decrease in the gyromagnetic ratio, and enhances the damping in $Y_3Fe_5O_{12}$. Such TSS-induced changes become more pronounced as the temperature decreases from 300 K to 50 K. These results suggest a completely new approach for control of magnetism in magnetic thin films.

Topological insulators can be characterized by a gap in the bulk band structure and a linear Dirac cone in the bulk gap for the surface states. When a topological insulator (TI) is interfaced with or in proximity to a magnetic insulator (MI) with perpendicular magnetization, the magnetic ordering in the MI can break the time-reversal symmetry of and open a gap at the Dirac point of the topological surface states (TSS) at the MI/TI interface. This can lead to formation of a quantum anomalous Hall insulator and an axion insulator when the Fermi level is within the gap opened by the MI;[1,2,3] when the Fermi level is not in the gap, the conventional anomalous Hall effect[4,5,6,7] and the topological Hall effect [8] are observed. Further, the proximity to a MI may also induce magnetic moments[9,10,11,12,13] and negative magnetoresistance[14] in the TI. These effects are intriguing fundamental phenomena and yet of great technological importance and have therefore attracted considerable interests in recent years.

In contrast, the inverse effect of the TSS in the TI on the magnetism in the MI remain largely unexplored. Theory indicates several interesting phenomena in this context: the TSS in a TI film can induce a perpendicular magnetic anisotropy (PMA) in an adjacent MI film,[15,16,17] lead to a change in the gyromagnetic ratio ($\gamma$) in the MI,[17] or produce an extra damping in the MI that should scale with the electron relaxation time.[17] So far, experiments have neither observed the formation of PMA nor a change in $\gamma$ induced by TSS. Recent experiments have demonstrated the enhancement of damping in MI/TI structures, but its relationship with the relaxation time has not been explored.[18,19,20]

This letter reports on the effects of the TSS on the magnetism in a bi-layered structure that consists of a topological insulator $Bi_2Se_3$ film grown on the top of a magnetic insulator $Y_3Fe_5O_{12}$ (YIG) film. Ferromagnetic resonance (FMR) in such bilayers was measured as a function of temperature ($T$), microwave frequency, and field angle. The analysis of the FMR data yields four important results. First, the TSS produce a PMA in the YIG film, as expected theoretically.[15,16,17] Second, the TSS results in a decrease of ~1% in the absolute gyromagnetic ratio ($|\gamma|$) of the YIG film. Third, the TSS significantly enhance the damping in the YIG, as predicted;[17] the resulting damping is one order of magnitude larger than the intrinsic damping. Finally, the TSS-produced effects become stronger when $T$ is decreased from 300 K to 50 K. These results can be understood in terms of the spatial penetration of the TSS into the YIG and the corresponding enhancement of spin-orbit coupling in the YIG.[16] One can speculate that the stronger effect at low $T$ is likely associated with the conductivity enhancement of the TSS at low $T$.[6] Control measurements clearly confirm that the measured changes of the magnetic properties of the YIG are not due to permanent changes in the YIG film quality created by the overgrowth of $Bi_2Se_3$. Rather, these changes are indeed due to the TSS. These results

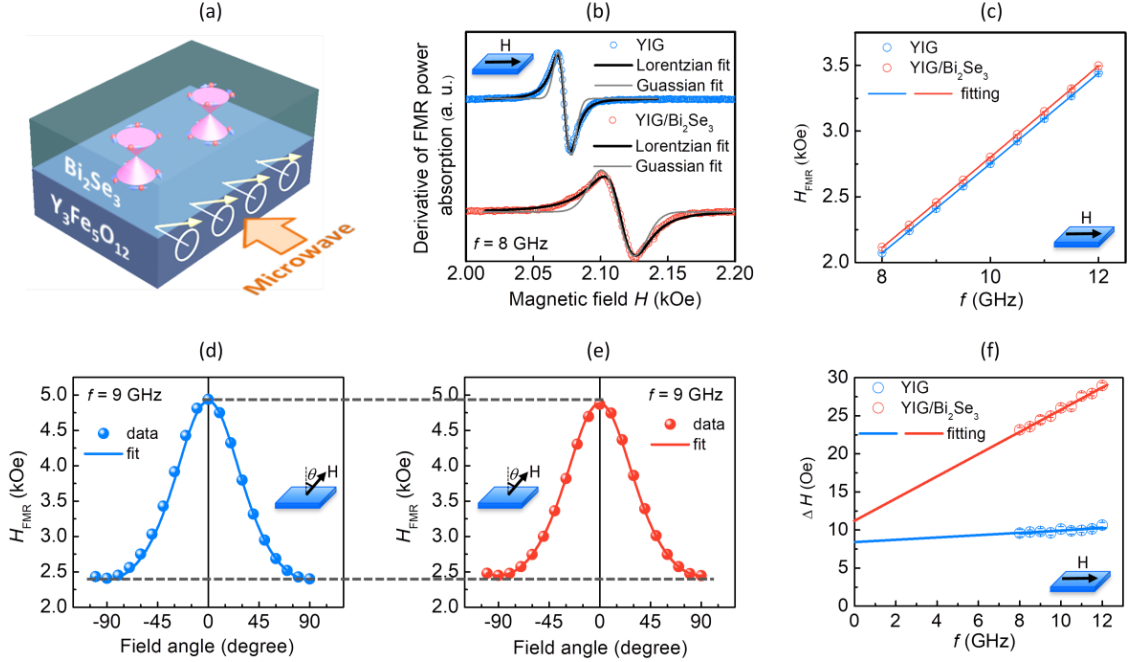

FIG. 1. Modulation of ferromagnetic resonance (FMR) properties of a 15-nm YIG film by the TSS in an adjacent 8-nm $Bi_2Se_3$ film. (a) Schematic of a YIG/$Bi_2Se_3$ bilayer. (b) FMR profiles measured on the YIG film before (blue) and after (red) the growth of the $Bi_2Se_3$ film on top. (c) and (f) show the FMR field $H_{FMR}$ and linewidth $\Delta H$, respectively, measured at different frequencies (f) for the YIG film before (blue) and after (red) the $Bi_2Se_3$ growth. (d) and (e) present $H_{FMR}$ measured as a function of the field angle $\theta$ on the YIG film before and after, respectively, the $Bi_2Se_3$ growth.

suggest a new way for control of magnetism in magnetic thin films.

Figure 1 presents the main data of this work. Figure 1(a) illustrates the experimental configuration. Figures 1(b)-1(f) present the FMR data measured at room temperature; the blue and red points show the data for the 15-nm-thick YIG film before and after, respectively, the growth of an 8-nm-thick $Bi_2Se_3$ film on the top. The details about the growth and characterization of these films are provided in the Supplementary Materials.[21] The TSS in the discussions below refer to the topological electronic states at the YIG/ $Bi_2Se_3$ interface, not the top surface of the $Bi_2Se_3$ film.

Figure 1(b) shows representative FMR profiles measured at a frequency (f) of 8 GHz in an in-plane field (H). The circles show the data; the black and gray curves show fits to the derivatives of a Lorentzian function and a Gaussian function, respectively. Three results are evident in Fig. 1(b). First, the growth of the $Bi_2Se_3$ film leads to a notable shift of the resonance to a higher field, which is due to TSS-produced PMA, as discussed shortly. Second, the $Bi_2Se_3$ growth results in linewidth broadening, a consequence of damping enhancement due to the TSS, as explained below. Third, the Lorentzian fit is better than the Gaussian fit. This indicates that the film inhomogeneity contribution to the FMR linewidth is small. The FMR field ($H_{FMR}$) and linewidth ($\Delta H$) data presented below are from the Lorentzian fitting; $\Delta H$ represents the peak-to-peak linewidth.

To confirm the above-described $H_{FMR}$ shift and $\Delta H$ enhancement, FMR measurements were repeated at different frequencies. Figure 1(c) presents the $H_{FMR}$ data, while the $\Delta H$ data are shown in Fig. 1(f). The lines in Fig. 1(c) show the fits to

$$f = |\gamma|\sqrt{H_{FMR}(H_{FMR} + 4\pi M_{eff})} \qquad (1)$$

where $4\pi M_{eff}$ denotes the difference between the saturation induction $4\pi M_s$ and anisotropy field $H_a$ of the YIG film, namely, $4\pi M_{eff} = 4\pi M_s - H_a$. $H_a > 0$ and $H_a < 0$ correspond to the presence of a PMA and an easy-plane anisotropy, respectively. The fitting of the data from the bare YIG film yields $|\gamma| = 2829 \pm 2$ MHz/kOe and $4\pi M_{eff} = 1785.5 \pm 6.7$ G. The $|\gamma|$ value is close to the standard value. The $4\pi M_{eff}$ value is close to the $4\pi M_s$ value of bulk YIG (1750 G),[22] indicating that the anisotropy in the YIG film is very weak.

The fitting of the data from the YIG/Bi$_2$Se$_3$ sample, however, yields different values. Specifically, one has $|\gamma| = 2807 \pm 2$ MHz/kOe, which is 22 MHz/kOe or 0.78% smaller than that of the bare YIG sample. This change is an order of magnitude larger than the error bars and therefore represents a real change. On the other hand, $4\pi M_{eff}$ is equal to $1724.7 \pm 7.7$ G, which is 60.8 G lower than that in the bare YIG sample. This reduction indicates either a decrease in $4\pi M_s$ of 60.8 G or the presence of a PMA with $H_a = 60.8$ Oe. Previous work has shown that in a MI/TI bilayer, the TSS should not result in a change in $4\pi M_s$.[16] Further, after the removal of the Bi$_2$Se$_3$ film the $4\pi M_{eff}$ value returns to the value of the bare YIG sample, as shown below. These facts suggest that the reduction of $4\pi M_{eff}$ is not a trivial consequence of a change in $4\pi M_s$ created simply by the growth of a Bi$_2$Se$_3$ film on the top of the YIG film. Thus, one can conclude that the reduction of $4\pi M_{eff}$ results from the presence of a PMA due to the TSS; such a TSS-produced PMA has been predicted by three separate studies but has never been observed experimentally.[15,16,17] Note that another possible reason for the $4\pi M_{eff}$ reduction is the dynamic separation of interface moments from the bulk moments in the YIG thin film, but this seems to be unlikely because the FMR profiles consist of well-defined single peaks with expected Lorentzian shape.

The data in Figs. 1(b) and 1(c) were measured with in-plane fields. To further confirm the above results about TSS-produced $|\gamma|$ reduction and PMA formation, FMR measurements on the same samples were also performed as a function of the field angle ($\theta$) relative to the film normal. Figures 1(d) and 1(e) present the $H_{FMR}$ vs. $\theta$ data for the bare YIG sample and the YIG/Bi$_2$Se$_3$ sample, respectively. As indicated by the horizontal dashed lines, the growth of the Bi$_2$Se$_3$ film leads to changes in the $H_{FMR}$ vs. $\theta$ response. To describe such changes more quantitatively, the data were fitted to[23]

$$\left(\frac{f}{|\gamma|}\right)^2 = [H_{FMR}\cos(\theta - \phi) - 4\pi M_{eff}\cos(2\phi)] \cdot [H_{FMR}\cos(\theta - \phi) - 4\pi M_{eff}\cos^2(\phi)] \quad (2)$$

where $\phi$ is the angle of the equilibrium magnetization relative to the film normal that can be found according to

$$H_{FMR}\sin(\theta - \phi) + \frac{1}{2}4\pi M_{eff}\sin(2\phi) = 0 \quad (3)$$

The fitting yields $|\gamma| = 2855 \pm 1$ MHz/kOe for the YIG sample and $|\gamma| = 2826 \pm 1$ MHz/kOe for the YIG/Bi$_2$Se$_3$ sample, which indicate a decrease of 1.02% in $|\gamma|$ due to the Bi$_2$Se$_3$ growth. The fitting also yields $4\pi M_{eff} = 1764 \pm 2$ G for the YIG sample and $4\pi M_{eff} = 1697 \pm 1$ G for the YIG/Bi$_2$Se$_3$ sample, which indicate the presence of a PMA field of ~67 Oe after the Bi$_2$Se$_3$ growth. These results support those from the $f$-dependent measurements.

In addition to modifying $|\gamma|$ and producing a PMA, the TSS also lead to an enhanced damping in the YIG, as indicated by the $\Delta H$ data in Fig. 1(f). In the figure, the circles show the $\Delta H$ data, while the lines show a fit to

$$\Delta H = \frac{2\alpha_{eff}}{\sqrt{3}|\gamma|}f + \Delta H_0 \quad (4)$$

where $\alpha_{eff}$ is the effective damping constant and $\Delta H_0$ denotes the inhomogeneity line broadening. The data indicate that the Bi$_2$Se$_3$ growth gives rise to a rather significant increase in $\Delta H$ of the YIG. For example, $\Delta H$ at $f$=10 GHz increased from 10.1±0.3 Oe for the bare YIG film to 26.0±0.3 Oe for the YIG/Bi$_2$Se$_3$ bilayer. The data also indicate that the increase of $\Delta H$ is mostly due to the increase of the slope of the $\Delta H$ vs. $f$ response, namely, due to a rise in $\alpha_{eff}$, rather than due to an increase in $\Delta H_0$. These results, together with the facts that (i) the Lorentzian function fits better the FMR profiles than the Gaussian function and (ii) after the removal of the Bi$_2$Se$_3$ film $\Delta H$ returns to the initial value for the bare YIG film, evidently indicate that the Bi$_2$Se$_3$ film-caused $\Delta H$ increase is not due to inhomogeneity line broadening, but is rather due to an enhancement in $\alpha_{eff}$.

The fitting in Fig. 1(f) yields $\alpha_{eff}$=(4.84±1.31)×10$^{-4}$ for the YIG film. This value is one or two orders of magnitude smaller than that in metallic films, indicating the high quality of the YIG. For the YIG/Bi$_2$Se$_3$ sample, however, the fitting yields $\alpha_{eff}$=(35.6±1.5)×10$^{-4}$, which is ~635% larger than that for the bare YIG film. If one defines $\Delta\alpha_{TSS}$ as the TSS-induced damping enhancement, one obtains $\Delta\alpha_{TSS}$≈30.8×10$^{-4}$.

The above-presented data show that the TSS in the Bi$_2$Se$_3$ can produce a PMA, result in a decrease in $|\gamma|$, and enhance $\alpha_{eff}$ in the YIG. These results may be attributed to the extension of the TSS into a few atomic layers of the YIG film near the interface, as discussed previously for EuS/Bi$_2$Se$_3$ bilayers where EuS is also a magnetic insulator.[16] Due to the spin-momentum locking of the TSS, such spatial penetration enhances spin-orbit coupling (SOC) in the YIG. Theory shows that this enhanced SOC can give rise to a PMA, as the magnetic anisotropy energy in a material is fundamentally related to the SOC of the material. On the other hand, the enhanced SOC can lead to a nonzero orbital moment in the YIG and thereby result in a change in the Landé $g$ factor. Note that $g$ depends on the ratio of the orbital moment ($\mu_L$) to the spin moment ($\mu_S$). In the absence of SOC, $\mu_L$ is usually quenched for any crystal structures with cubic symmetry, and one has $\mu_L/\mu_S=0$ and $g=2$. In the presence of SOC, however, $\mu_L$ is no longer quenched and $g$ is no longer equal to 2.[24,25] As $|\gamma|$ is related to $g$ through $|\gamma|=\frac{g\mu_B}{\hbar}$, a change in $g$ results in a change in $|\gamma|$. Here, $\mu_B$ is the Bohr magneton, and $\hbar$ is the reduced Planck's constant.

Further, the TSS extension at the interface also results in strong coupling between the magnetization in the YIG and the conduction electrons in the Bi$_2$Se$_3$; this interfacial coupling produces additional damping,[17] namely, $\Delta\alpha_{TSS}$, for the magnetization dynamics in the YIG, via a process similar to intra-band magnon-electron scattering in ferromagnets. The intra-band scattering involves the breathing of the Fermi surface driven by magnetization precession and the re-population of electrons to the breathing Fermi surface; the lag in the electron re-population causes a damping to the precession.[26,27,28]

With the above elucidation, one would naturally expect that the effects should be stronger at low $T$ because the TSS in Bi$_2$Se$_3$ is more pronounced at low $T$. It is known that, because the scattering of surface state electrons with phonons in Bi$_2$Se$_3$ becomes weaker as $T$ decreases, the conductivity of the TSS ($\sigma_{TSS}$) increases with a decrease in $T$, giving rise to more robust TSS at low $T$.[6] To check this expectation, FMR measurements on the exactly same samples were carried out as a function of $T$. Figure 2 shows the main data, which were measured with a 9.47-GHz cavity. Figure 2(a) presents the $H_{FMR}$ vs. $T$ data. For both the samples, $H_{FMR}$ decreases with a decrease in $T$. This results from the $T$-dependence of $4\pi M_s$. The data also show that $H_{FMR}$ for the YIG/Bi$_2$Se$_3$ sample is higher than that for the bare YIG film over the entire $T$ range, as for the above-presented room-temperature data. Figure 2(b) shows the difference between the two $H_{FMR}$ fields ($\delta H_{FMR}$). The data show a clear overall trend, namely, that $\delta H_{FMR}$ increases with a decrease in $T$. This is consistent with the expectation; according to Eq. (1), a higher field is needed to satisfy the FMR condition if $H_a$ is increased and $|\gamma|$ is reduced due to a decrease in $T$. Figure 2(c) presents the corresponding $\Delta H$ data. One can see that the $Bi_2Se_3$-induced $\Delta H$ enhancement becomes much more significant with a decrease in $T$. Since the $Bi_2Se_3$-induced $\Delta H$ enhancement is mainly due to $\Delta\alpha_{TSS}$, as justified above, one can evaluate $\Delta\alpha_{TSS}$ at different $T$ using the data in Fig. 2(c). The resultant $\Delta\alpha_{TSS}$ values are given in Fig. 2(d). It is evident that $\Delta\alpha_{TSS}$ increases with a decrease in $T$, as expected.

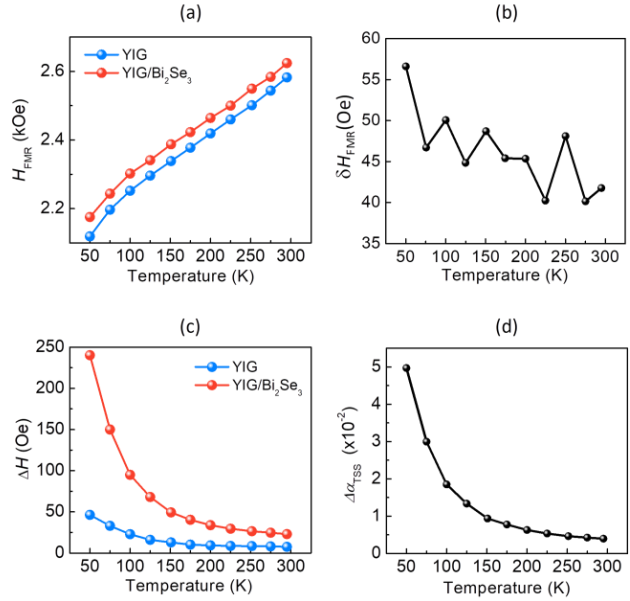

FIG. 2. Temperature ($T$) dependence of TSS-induced modulation of FMR properties in a YIG/Bi$_2$Se$_3$ bilayer. (a) and (c) show FMR field $H_{FMR}$ and linewidth $\Delta H$, respectively, as a function of $T$ of a 15-nm YIG film before and after the growth of an 8-nm Bi$_2$Se$_3$ film on the top. (b) Difference of $H_{FMR}$ before and after the Bi$_2$Se$_3$ growth, as a function of $T$. (d) TSS-produced damping enhancement $\Delta\alpha_{TSS}$ as a function of $T$. The FMR data were measured at 9.47 GHz under an in-plane field.

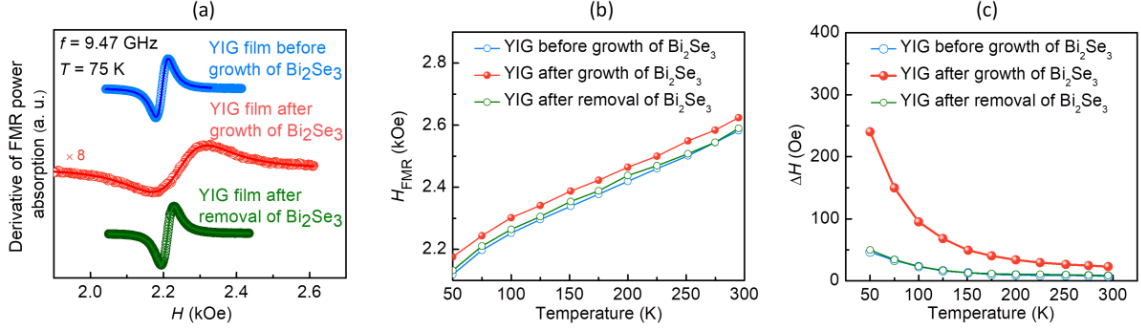

FIG. 3. Comparison of FMR properties of a 15-nm YIG film before and after the growth of an 8-nm $Bi_2Se_3$ film on the top, and after the removal of the $Bi_2Se_3$ film. (a) Comparison of FMR profiles. (b) Comparison of the FMR field $H_{FMR}$ vs. $T$ responses. (c) Comparison of the FMR linewidth $\Delta H$ vs. $T$ responses. The FMR data were all measured at 9.47 GHz under an in-plane field.

In spite of the above-presented consistency, there is a possibility that the $Bi_2Se_3$-induced changes of the YIG properties presented in Figs. 1 and 2 are not due to the TSS, but rather due to permanent changes in the YIG properties caused by the overgrowth of $Bi_2Se_3$. To rule out this possibility, control measurements were conducted that compare the FMR properties of the YIG film prior to the $Bi_2Se_3$ growth, after the $Bi_2Se_3$ growth, and after the removal of the $Bi_2Se_3$ film. The $Bi_2Se_3$ removal was realized through annealing of the sample in Ar at 350 °C for 2 min.

Figure 3 presents the main data. The data in Fig. 3(a) show that the growth of a $Bi_2Se_3$ film results in a notable shift of the resonance to a higher field and a substantial broadening of the resonance; after the removal, however, the resonance changes back to that of the YIG film before the $Bi_2Se_3$ growth. Figures 3(b) and 3(c) compare the $H_{FMR}$ and $\Delta H$ data, respectively. The data show that after the removal of the $Bi_2Se_3$, the FMR properties of the YIG are notably different from those of the YIG film with a $Bi_2Se_3$ capping layer but are very similar to those of the YIG film before the growth of $Bi_2Se_3$ over the entire $T$ range. These results prove that the changes of the magnetic properties presented in Figs. 1 and 2 are not due to the $Bi_2Se_3$ growth-produced "permanent" changes in the YIG properties but are rather associated with the TSS in the $Bi_2Se_3$ film.

Additional control measurements were performed on the bi-layered samples where the 8-nm $Bi_2Se_3$ capping layer was replaced by an 8-nm Au film, which is a heavy metal, or an 8-nm $(Bi_{0.75}In_{0.25})_2Se_3$ film,[29,30,31,32] which is a topologically trivial insulator. The measurements show that the changes of $|\gamma|$ and $\alpha_{eff}$ due to the growth of the Au or $(Bi_{0.75}In_{0.25})_2Se_3$ film are significantly smaller than the above-present changes, and the growth of Au or $(Bi_{0.75}In_{0.25})_2Se_3$ did not result in a PMA in the YIG, as discussed in the Supplementary Materials. As $(Bi_{0.75}In_{0.25})_2Se_3$ has a crystal structure and a bulk band structure very similar to those in $Bi_2Se_3$, the comparison of the results of YIG/$(Bi_{0.75}In_{0.25})_2Se_3$ and YIG/$Bi_2Se_3$ suggests that the changes of the YIG magnetic properties due to the $Bi_2Se_3$ growth are unlikely associated with the interfacial hybridization of electronic states below the Fermi level.

Several pertinent remarks should be made about the above results. (1) It is known that the FMR method represents an effective technique for measuring $|\gamma|$ and has been previously used to study how $|\gamma|$ in ferromagnetic films varies with the film thickness.[24,25] (2) Previous theoretical work indicates that $\Delta\alpha_{TSS}$ should scale with the electron relaxation time ($\tau$).[17] Since $\tau$ increases with a decrease in $T$, one would expect $\Delta\alpha_{TSS}$ to also increase with a decrease in $T$. This means that the $T$ dependence of $\Delta\alpha_{TSS}$ in Fig. 2(d) may have two origins, the $T$ dependence of $\sigma_{TSS}$ and the $T$ dependence of $\tau$. This likely explains why $\Delta\alpha_{TSS}$ exhibits a much stronger $T$ dependence than $\delta H_{FMR}$. (3) $\Delta\alpha_{TSS}$ had also been observed recently in similar bilayers,[18,19] but the $T$ dependence of $\Delta\alpha_{TSS}$ in Fig. 2(d) was not observed. Further, in-plane anisotropy, rather than PMA, was observed in those bilayers.[18,19] (4) One can expect that the TSS-produced effects should be stronger for a thinner YIG film but weaker for a thicker film, due to the nature of the TSS penetration at the interface. Indeed, weaker effects were observed in a YIG(30 nm)/Bi$_2$Se$_3$(8 nm) sample (see the Supplementary Materials). FMR measurements on bilayers with thinner YIG films were also performed, but the data were relatively noisy and did not yield reliable results. (5) As discussed in the Supplementary Materials, the transport and angle-resolved photoemission measurements on the Bi$_2$Se$_3$ film indicate that the Fermi level is in the bulk conduction band. Future work is of great interest that tunes the Fermi level via voltage gating or composition control and thereby study how the Fermi level influences the TSS-produced effects.[20] (6) The Bi$_2$Se$_3$ films in this work are 8 nm thick, and films thinner than 8 nm may host strong coupling between the states on the opposite surfaces.[33,34] Future work is also of interest that explore how such films modify magnetism in adjacent MI films.

Acknowledgement: This work was supported by the U.S. Department of Energy, Office of Science, Basic Energy Sciences (DE-SC0018994). The fabrication and characterization of the measurement samples were supported mainly by the U.S. National Science Foundation (EFMA-1641989; ECCS-1915849). Work at PSU was supported by the Penn State Two-Dimensional Crystal Consortium-Materials Innovation Platform (2DCC-MIP) funded by the U.S. National Science Foundation (DMR-1539916).

†Corresponding author. mwu@colostate.edu